\documentclass[a4paper,aps,prl,twocolumn,showpacs]{revtex4}

\draft 
\usepackage{graphicx}
\usepackage{amsmath}
\usepackage{amssymb}
\usepackage{natbib}
\usepackage{psfrag}
\usepackage{wasysym}
\usepackage{pifont}
\usepackage{epstopdf}
\usepackage{color}
\usepackage{url}
\usepackage{dcolumn}
\usepackage{bm}
\usepackage[normalem]{ulem}

\newcommand \beq{\begin{equation}}
\newcommand \eeq{\end{equation}}

\begin{document}

\title{Frictional Effects in Biomimetic Scales Engagement}

\author{Ranajay Ghosh$^1$, Hamid Ebrahimi$^1$, and Ashkan Vaziri$^1$}\email{vaziri@coe.neu.edu}

\affiliation{$^1$ Department of Mechanical and Industrial Engineering, Northeastern University, Boston, MA, 02115, USA\\}

\date{August 12,  2015}

\begin{abstract}
\noindent Scales engagement can contribute significantly to nonlinear bending behavior of elastic substrates with rigid biomimetic scales. In this letter, we investigate the role of friction in modulating the nonlinearity that arises due to self-contact of scales through an analytical investigation. We model the friction as dry Coulomb type friction between rigid links and the substrate is taken to be linear elastic. Our results reveal that frictional effects give rise to two possible locking mechanisms, namely static friction lock and kinetic friction lock. These locks arise due to a combination of interfacial behavior and geometry.  In addition to these extremes, the frictional behavior is found to increase stiffness of the structure. This dual nature of friction which influences both system operation and its terminal limit results in the maximum relative frictional work to lie at intermediate friction coefficients and not at the extremes of frictional limits.
\end{abstract}


\maketitle

\begin{figure}[h]
\centering
\includegraphics[width=1\columnwidth]{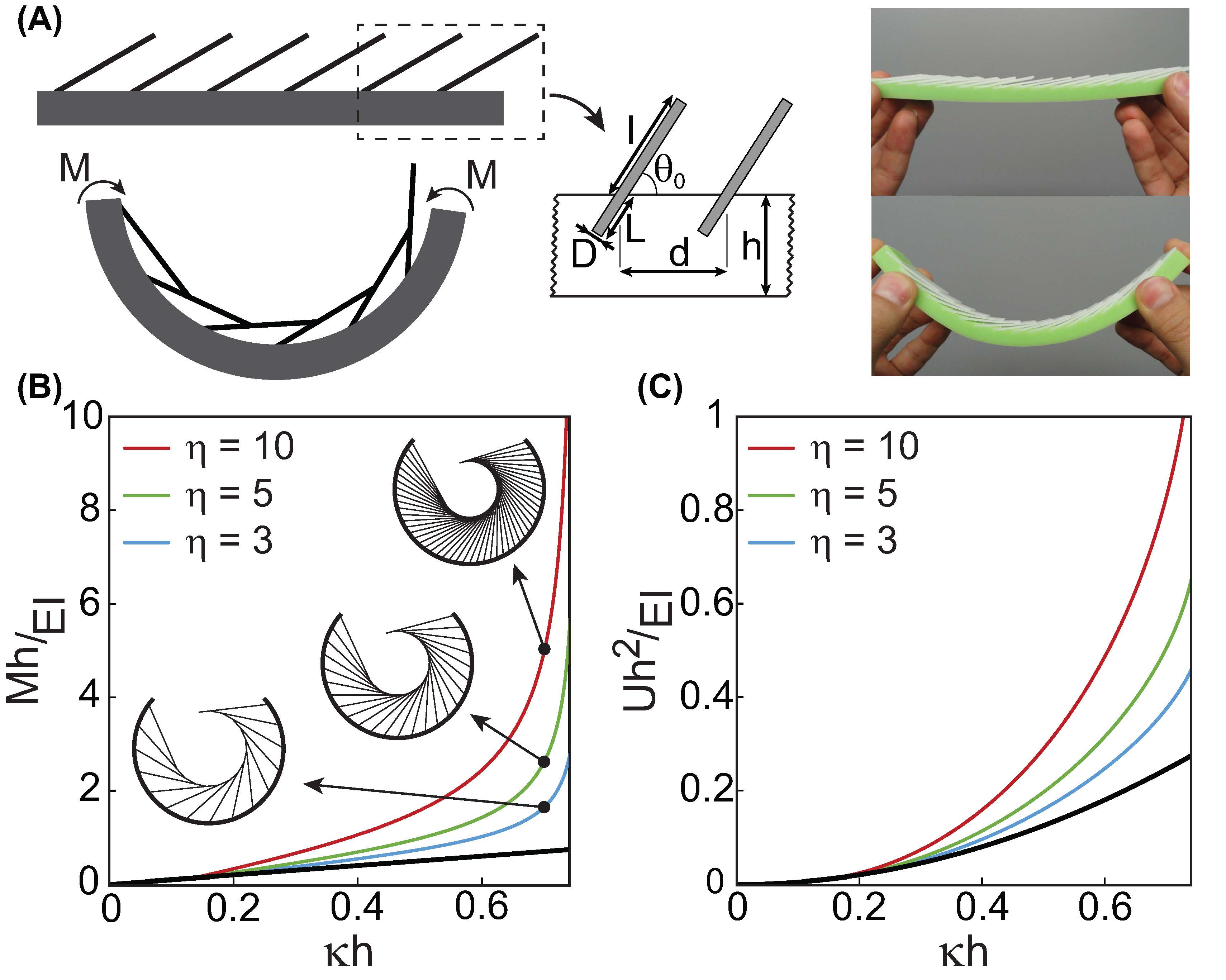}
\caption{(A) (left)A schematic of biomimetic scale system studied in the paper along with the (inset) geometrical dimensions (right) A 3$-$D printed material specimen of the proposed biomimetic structure (B) Non-dimensionalized moment-curvature response for biomimetic systems of unit width with various overlap ratios ($\eta$) and  scales at an initial angle of 10 degrees. Scale geometrical ratios assumed as $L/D=15$,   $h/L=5$ and $d/D>5$. The thick black curve indicates virgin beam with no scales with (inset) schematic of deflection shapes plotted at $\kappa h=0.7$. (C) Non-dimensionalized energy density-curvature response for the same system.}
\label{fig:1}
\end{figure}

\noindent Dermal scales can provide considerable mechanical \cite{R1, R2, R3} and multifunctional advantages \cite{R4, R5, R6, R7, R8} without adding significant weight to organisms making them ubiquitous in the animal kingdom. These properties along with their relative simplicity make them attractive candidates for adoption into synthetic materials development paradigms in order address the challenges that arise from balancing weight, mechanical performance and functionality. To this end, elementary structures mimicking the natural scale-systems can be fabricated through partially embedding plate like surface appendages in a soft substrate \cite{R9, R10, R11, R12, R13}. The analysis of these systems which exhibit rich mechanical behavior indicate that the primary source of complexity is the engagement of scales with each other \cite{R9, R10, R11, R12, R13}. This mechanical enhancement, tailorable through a straightforward variation of geometry, could be leveraged for material systems which require tunability of elastic behavior and energetic landscape. For a compliant beam like substrate, this scale engagement , which can occur due to change of curvature of the parent beam, induces remarkable nonlinearity in bending through a self-contact mechanism \cite{R13}. For rigid scales, this self-contact mechanism was found to ultimately lead towards a kinematically dictated locked configuration. In the idealized system shown in Fig.~\ref{fig:1}(A) where scales deformation is assumed to be negligible with respect to the elastic substrate, the extent of nonlinearity in system response depends on the ratio of scale length to scale spacing ($\eta=l/d$), scale embedded length ($L$) , scale thickness ($D$) and the elastic modulus of the substrate ($E$)\cite{R13}. For instance, variation in $\eta$ can be used to significantly influence both bending stiffness and elastic energy absorption variation with curvature, Fig.~\ref{fig:1}(B-C). In this context, the effect of friction becomes important for furthering the understanding of these systems since the origin of nonlinearity stems from scale contacts.

\begin{figure}[h]
\centering
\includegraphics[width=1\columnwidth]{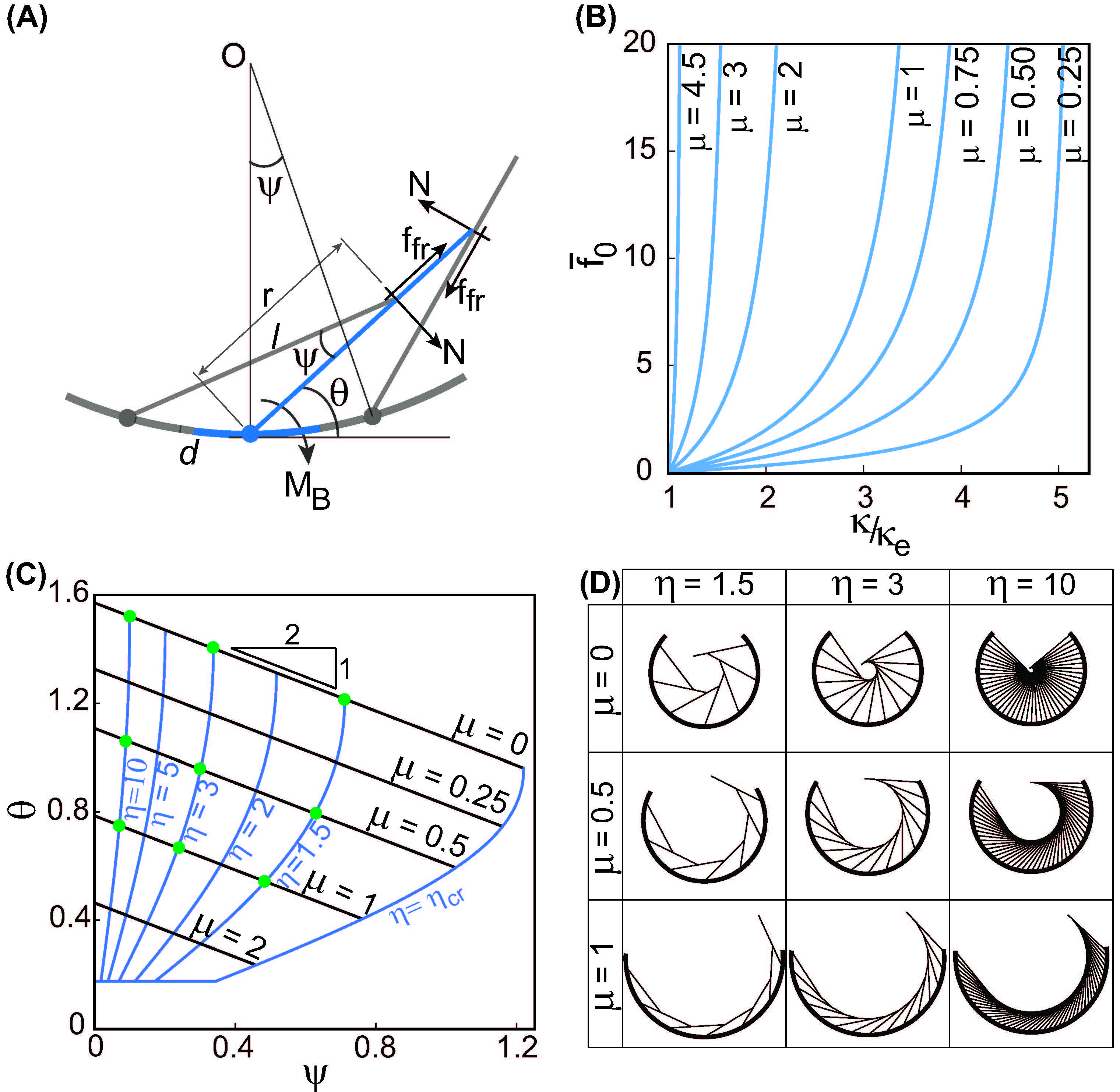}
\caption{(A) Scale representative volume element (RVE) geometry along with free body diagram of an individual scale (B) Non-dimensionalized ($\kappa_e$ is the engagement curvature) frictional force for various friction coefficients. It is clear that friction forces approach singularity near the frictional locking configuration. In this simulation $\eta=6$ and the initial inclination $\theta_0=10^o$ (C) Phase map representation of the biomimetic system delineates the role of friction in mechanical behavior. The map shows three distinct regimes of performance – linear, kinematically determined nonlinear and a frictionally determined locking boundary (D) A schematic visual representation of the locking state indexed by friction coefficient $μ$ and $\eta$.}
\label{fig:2}
\vspace{-2 em}
\end{figure}

In this letter we investigate the role of friction in modulating the influence of biomimetic scales on a beam in pure bending. We assume rigid scales, linear elastic substrate and Coulomb type friction in order to isolate the role of critical geometrical and interfacial parameters in fostering mechanical nonlinearity as distinct from any traditional material sources. The nature of the nonlinearity becomes apparent if the system is simplified and periodic self-similarity of contacting scales is assumed. This periodicity allows us to isolate a fundamental representative volume element (RVE) for further analysing the system, Fig.~\ref{fig:2}(A). A geometrical scrutiny of the RVE results in the following nonlinear relationship between the scale rotation angle and the substrate rotation, Fig.~\ref{fig:2}(A)\cite{R13}:

\beq
\eta\psi \cos\psi/2 - \sin(\theta+\psi/2)=0
\label{eqn:eq1}
\eeq

where $\psi$ is the angular rotation of the substrate of RVE with respect to the instantaneous center of curvature of the beam, Fig.~\ref{fig:2}(A) and which is directly related to instantaneous curvature $\kappa$ of the substrate as $\psi=\kappa d$.  We also impose $\theta\leq\pi/2$ since we are currently considering only the “right chirality” inclination and $\psi\leq\pi$ , the geometrically permissible range. Note that this relationship will not be satisfied if the $\eta$ is too low, indicating no engagement. In general this limiting ratio is given by $\eta_{cr}=\sin\theta_0/\theta_0$ where $\theta_0$ is the initial scale inclination [13].  For values of $\eta>\eta_{cr}$, the structure starts to deviate from the linear behavior after scale engagement according to Eq.~\ref{eqn:eq1}. The equation predicts that as the scales slide sufficiently from their initial inclination, they approach a limiting locking configuration given by $\theta_{lock}$ and a corresponding limiting curvature given by $\psi_{lock}$ where sliding is impossible. These terminal angles can be shown to be inter-related by the relationship $\theta_{lock}+\psi_{lock}/2=\pi/2$ [13].

In order to understand the role of friction in influencing this behavior, we first look at the free body diagram of the RVE post-engagement, Fig.~\ref{fig:2}(A). Note that the frictional force $f_{fr}$ depends on the direction of relative motion between scales. This can be discerned \textit{a priori} from the kinematic relationship derived earlier, Eq.~\ref{eqn:eq1} whose solution has two branches. Both branches meet at the locking configuration ($\theta_{lock}$, $\psi_{lock}$) and depending on the initial value of $\theta=\theta_0$, one of them would be physically relevant. We call the branch, which corresponds to $\theta_0<\theta_{lock}$, the low angle branch where both scale angle and curvature increase and the other one corresponding to $\theta_0>\theta_{lock}$, the high angle branch where the scale angle decreases with curvature. Thus, we can express the friction force  in terms of magnitude $f_0$ as $f_{fr}=f_0 sgn(\theta_{lock}-\theta_0)$ and obtain the following expression from the balance of moments at the base of a single scale, Fig.~\ref{fig:2}(A)

\beq
N(l\cos\psi-r) - f_0l~sgn(\theta_{lock}-\theta_0)\sin\psi=K_B(\theta-\theta_0)
\label{eqn:eq2}
\eeq

where, $r$ is the distance from the base to the point of contact between the scales, Fig.~\ref{fig:2}(A), $K_B=0.66~E D^2 (L/D)^{1.75}$ \cite{R13} is the rotational stiffness measuring the interaction (assumed linear) of the rigid scale with the elastic material into which it is embedded and $N$ is the inter-scale normal reaction force.
The normal reaction force can itself be written in terms of $f_0$ and the coefficient of friction $\mu$ using the Coulumb friction law $f_0\leq\mu N$ with the equality holding in the kinetic friction regime. We use the same coefficient of friction for the static as well as kinetic regime although in practice, the static friction coefficient is slightly more than its kinetic counterpart. Thus we arrive at the following expression for the magnitude of the frictional contact force (see SM for derivation):

\vspace{-1 em}
\beq
\overline{f}_0=\frac{f_0l}{K_B}\leq \sin\beta \frac{\theta-\theta_0}{\cos(\psi+sgn(\theta_{lock}-\theta_0)\beta)-\overline{r}\cos\beta}
\label{eqn:eq3}
\eeq

where $\overline{r}=r/l$, $\beta=\tan^{-1}\mu$ (friction angle). Note that the equality is imposed as soon as sliding commences.

\begin{figure}[h]
\centering
\includegraphics[width=1\columnwidth]{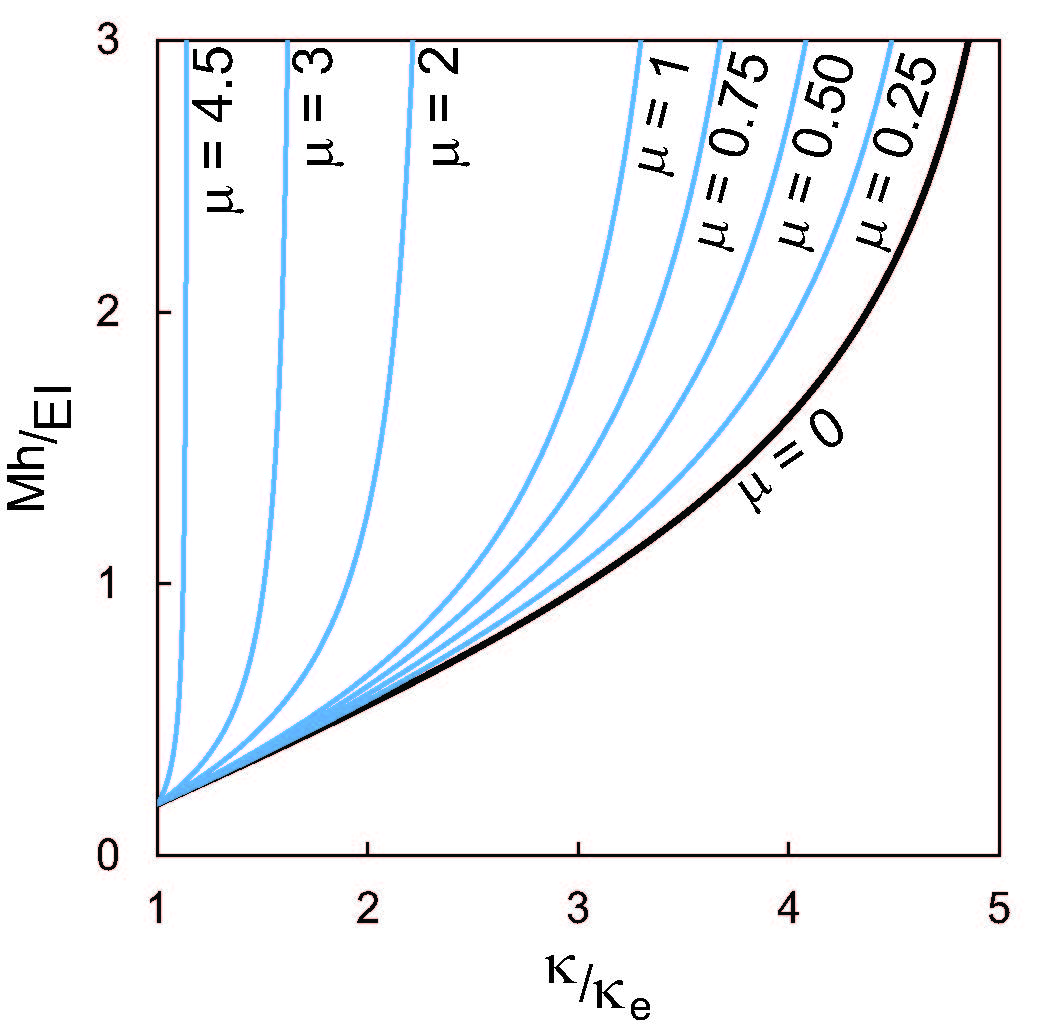}
\caption{Non-dimensionalized ($\kappa_e$ is the engagement curvature) post engagement moment-curvature relationships for various values of friction coefficient clearly showing the perceptible difference brought about by friction in effective bending stiffness of the biomimetic structure. For this simulation $\eta=6$ and the initial inclination $\theta_0=10^o$. Scale geometrical ratios assumed as $L/D=15$, $h/L=5$ and $d/D>5$.}
\label{fig:3}
\end{figure}

 The introduction of friction alters the purely kinematic nature of locking behavior described earlier. In this case, the locking mechanism is now governed by interfacial rather than purely kinematic variables. This can be ascertained from the frictional relationship, Eq.\ref{eqn:eq3}, which would yield, at a certain curvature, an infinite frictional force giving rise to another locking mechanism governed by friction. As the system approaches this locking region, motion arrest begins due to escalating frictional forces, Fig.~\ref{fig:2}(B), resulting in a limiting “kinetic friction lock” which would always be lesser than the theoretical frictionless locking curvature.

 The overall nonlinear behavior of this frictional system can be unified through a phase map spanned by $\theta-\psi$, which plots the dependence of scale rotation $\theta$ with substrate rotation $\psi$ for various values of $\eta$ based on Eqs.\ref{eqn:eq1} and \ref{eqn:eq3}, Fig.~\ref{fig:2}(C-D). In this map, the frictional locking configuration, which is governed by the frictional coefficient is represented by the kinetic frictional lock line in the phase space defined by $\theta_{lock}^{'}+(\psi_{lock}^{'})/2=\pi/2-sgn(\theta_{lock}-\theta_0)\beta$ where prime denotes frictional locking configuration. This expression quantifies the friction mediated advance of the locking envelope alluded to earlier.
 Moreover, the nature of Coulomb friction ensures that as long as the initial scale angle corresponds to the low angle branch, any initial angle greater than the frictional locking angle would cause an instantaneous post-engagement “static frictional lock” of the system. However, if the initial angle ensures the higher angle branch i.e. $\theta_0>\theta_{lock}$, then any initial angle lower than the frictional lock angle will result in the static frictional lock. Thus clearly there exists a “static friction lock band” in the phase space of width $2\beta$. Thus, a reference configuration defined by an initial angle and $\eta$ may cause the system to fall in this band and thereby rendered rigid immediately after engagement.  However, if the configuration lies outside this band, the scales will move till they reach the boundary of this zone. Alternatively, the critical static frictional locking coefficient $\mu'$ for a given initial angle and $\eta$ can also be evaluated from the frictional locking line, $\theta_0+(\psi'_{lock}(\eta,\mu))/2=\pi/2-sgn(\theta_{lock}-\theta_0)\tan^{-1}\mu$. Without loss of generality, taking the low angle branch, we can obtain the critical static frictional coefficient by solving the transcendental equation  $\mu'=\cot(\theta_0+\psi(\mu',\theta_0))$. This shows that, for a given reference configuration, a frictional coefficient greater than $\mu'$ would result in an instantaneous post-engagement locking corresponding to static frictional lock. Interestingly, from Fig.~\ref{fig:2}(C) it becomes clear that static frictional locking would be harder to achieve in the higher $\eta$ and lower initial angles region. Thus closely packed and grazing (low angle) scales, widely found in nature seem to be particularly resistant to frictional locking.

Next, in order to understand the mechanics of this system we now derive the moment$-$curvature relationship of this structure. Assuming periodicity and Euler-Bernoulli beam behavior, we equate the work done due to bending to the total energy absorbed by the system and the frictional dissipation. This leads to the following variational energetic equation:

\begin{multline}
\int_{0}^{\kappa} M \,d\kappa^{'} = \frac{1}{2} E I \kappa ^2 +\Big[\frac{1}{2} \frac{1}{d}K_B(\theta-\theta_0)^2+ \\  \frac{1}{d}\int_{\kappa_e}^{\kappa} f_0~sgn(\theta_{lock}-\theta_0) \,dr\Big]~H(\kappa-\kappa_e)
\label{eqn:eq4}
\end{multline}

where $M$ is the applied moment,  $\kappa, \kappa_e$ are respectively the current and engagement curvatures of the beam, $I$ is the second moment of area of the beam cross-section and $H(.)$ is the Heaviside step function for tracking engagement. Note that although the frictional work contains a singular kernel, the work integral exists. We justify linear material behavior assuming relatively small strains in the substrate. Next, taking derivative of Eq.\ref{eqn:eq4} with respect to the current curvature yields the moment curvature relationship which can be evaluated numerically (See SM for derivation). We plot this relationship in a non-dimensional form for a system with scale geometry described in the introduction of the letter with various friction coefficients in the post-engagement regime to isolate frictional contribution, Fig.~\ref{fig:3}. This figure clearly shows that friction can perceptibly increase the bending stiffness of the structure and since it also controls the locking envelope, the increase can be made to occur at much lower curvatures.  Thus friction has a dual contribution to the moment-curvature relationship modulating both the magnitude and shape of the response.

\begin{figure}[h]
\centering
\includegraphics[width=1\columnwidth]{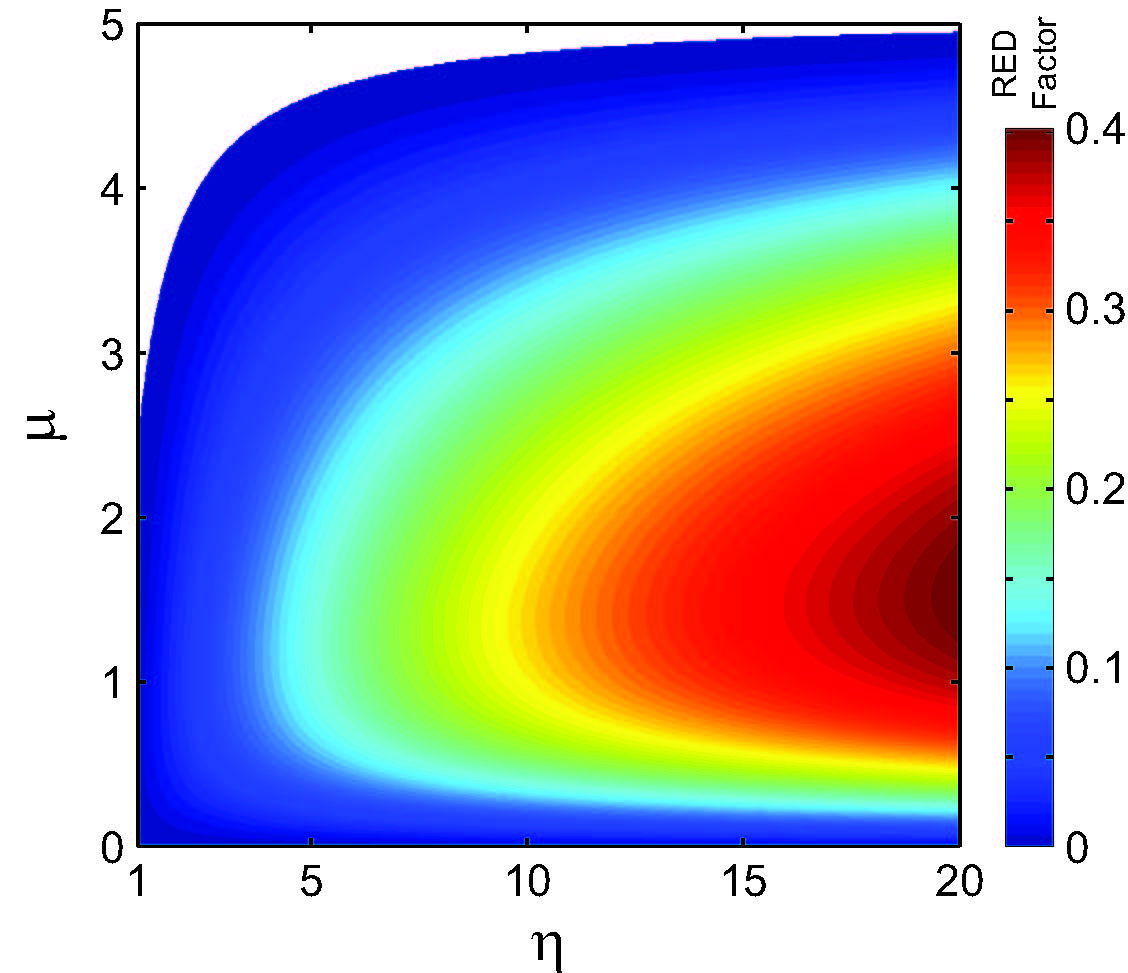}
\caption{Non-dimensional relative energy dissipation (RED) factor phase map spanned by the $\eta$ and frictional coefficient $\mu$. The dissipation landscape shows maxima at the interior away from the boundaries indicating dissipation maximization at intermediate frictional coefficients. Scale geometrical ratios assumed as $L/D=15$, $h/L=5$ and $d/D>5$.}
\label{fig:4}
\end{figure}

This dual contribution of the frictional forces i.e. both increasing bending stiffness and lowering the locking curvature can have significant implications on the maximum frictional work during bending. To this end we define the relative energy dissipation (RED) factor as the ratio of frictional work $w_{fr}$ (per unit length) to the work done on the system till locking $w_{sys}$ (per unit length),
\begin{equation}
RED = \frac{w_{fr}}{w_{sys}}
\label{eqn:eq5}
\end{equation}
where $w_{fr}=\frac{1}{d}\int_{\kappa_e}^{\kappa^{'}_{lock}}f_0~sgn(\theta_{lock}-\theta_0) \,dr$, $w_{sys}=w_{fr}+u_{el}$ with $u_{el}=\frac{1}{2}EI\kappa^2+\frac{1}{2}\frac{1}{d}K_B(\theta_{lock}^{'}-\theta_0)^2$, the elastic energy density of the system and the frictional locking curvature is $\kappa_{lock}^{'}=\psi_{lock}^{'}/d$. Note that RED is a function of $\eta$, friction coefficient $\mu$ and initial scale inclination $\theta_0$. However, fixing the initial angle $\theta_0=10^o$ for the current simulation allows us to obtain a dissipation phase plot which maps this non-dimensional dissipation factor through the system parameters $\eta$ and $\mu$, Fig.~\ref{fig:4}. This map shows the RED factor passes through maxima with increasing $\mu$ and then wanes for any given $\eta$. Thus unilaterally increasing $\mu$ does not necessarily improve the dissipation capacity of the system. This indicates that in the limits of low and high friction the relative contribution of frictional work diminishes but through different mechanisms $-$ lower overall moment contribution at low $\mu$ and through lowering of the locking limit for higher $\mu$, corresponding to the dual nature of the frictional force. Finally, note that this frictional mechanism does not stem from or give rise to any distinction between the loading and unloading paths unlike classical hysteresis.

In conclusion, we find that scale friction can advance the locking envelope and make high angle motion inaccessible. Interestingly, if friction coefficient is larger than a critical value for a given configuration, it can instantaneously arrest post-engagement motion, the so called static friction lock. Thus friction introduces a “lock band” in the system which can have useful technological implications.  Furthermore, we find that depending on the sought role of the system in a particular application, such as stiffness gain or energy attenuation, the corresponding frictional property desired may be different. Thus, our study demonstrates that in addition to scale geometry and substrate properties, engineering scale surfaces to modulate friction can provide an important avenue for designing and tuning biomimetic scales systems.
\par This report was made possible by the United States National Science Foundation's, Civil, Mechanical and Manufacturing Innovation Grant No. 1149750 and a NPRP award [NPRP 5-068-2-024] from the Qatar National Research Fund (a member of the Qatar Foundation). The statements herein are solely the responsibility of the authors.

\end{document}